\documentstyle  {article}
\begin{document}
\rightmargin -2.75cm
\textheight 23.0cm
\topmargin -0.5in
\baselineskip 16pt
\parskip 18pt
\parindent 30pt
\title{ \large \bf Flavour Democracy Calls for the Fourth Generation}

\author{Amitava Datta\\ Physics Department, Jadavpur University,\\
Calcutta - 700 032, India\\
 and\\
 International Centre for Theoretical Physics,\\
I 34100, Trieste, Italy}
\date{}
\pagestyle{empty}
\pagenumbering{arabic}
\baselineskip 24pt
\maketitle
\vspace{.2in}
\baselineskip 12pt
\vspace{0.5cm}
\pagestyle{empty}
\baselineskip 16pt
\parskip 16pt
\pagestyle{plain}
\begin{abstract}
It is argued with the  help of an illustrative model,
that the inter--species  hierarchy among
the  fermion masses and the quark mixing angles  can be accommodated
naturally in the standard model with (approximate) flavor  democracy provided
there are four families  of sequential quark--leptons with all members of
the fourth family having roughly equal masses. The special problem of
light neutrino masses (if any) and possible solutions are  also discussed.
\end{abstract}
\newpage
\def\singlespace {\smallskipamount=3.75pt plus1pt minus1pt
                  \medskipamount=7.5pt plus2pt minus2pt
                  \bigskipamount=15pt plus4pt minus4pt
                  \normalbaselineskip=15pt plus0pt minus0pt
                  \normallineskip=1pt
                  \normallineskiplimit=0pt
                  \jot=3.75pt
                  {\def\smallskip {\vskip\smallskipamount}}
                  {\def\medskip   {\vskip\medskipamount}}
                  {\def\bigskip   {\vskip\bigskipamount}}
                  {\setbox\strutbox=\hbox{\vrule
                    height10.5pt depth4.5pt width 0pt}}
                  \parskip 7.5pt
                  \normalbaselines}
\def\middlespace {\smallskipamount=5.625pt plus1.5pt minus1.5pt
                  \medskipamount=11.25pt plus3pt minus3pt
                  \bigskipamount=22.5pt plus6pt minus6pt
                  \normalbaselineskip=22.5pt plus0pt minus0pt
                  \normallineskip=1pt
                  \normallineskiplimit=0pt
                  \jot=5.625pt
                  {\def\smallskip {\vskip\smallskipamount}}
                  {\def\medskip   {\vskip\medskipamount}}
                  {\def\bigskip   {\vskip\bigskipamount}}
                  {\setbox\strutbox=\hbox{\vrule
                    height15.75pt depth6.75pt width 0pt}}
                  \parskip 11.25pt
                  \normalbaselines}
\def\doublespace {\smallskipamount=7.5pt plus2pt minus2pt
                  \medskipamount=15pt plus4pt minus4pt
                  \bigskipamount=30pt plus8pt minus8pt
                  \normalbaselineskip=30pt plus0pt minus0pt
                  \normallineskip=2pt
                  \normallineskiplimit=0pt
                  \jot=7.5pt
                  {\def\smallskip {\vskip\smallskipamount}}
                  {\def\medskip   {\vskip\medskipamount}}
                  {\def\bigskip   {\vskip\bigskipamount}}
                  {\setbox\strutbox=\hbox{\vrule
                    height21.0pt depth9.0pt width 0pt}}
                  \parskip 15.0pt
                  \normalbaselines}
\parskip .1in

\singlespace
The neutrino counting at LEP has established with high
statistical significance that the number of light
neutrinos ($m_{\nu} <<  m_{Z} /2$ )
is three \cite
{lep90}. This has lead to the widespread belief
that at least within the framework of the Glashow--Salam--Weinberg
minimal standard model ( MSM ) with
sequential quark--lepton families, the number of such families has
also been determined to be three\cite{comment,paschos}.
This belief hinges on the  notion that
a heavy sequential Dirac neutrino
($m_{\nu} >>  m_{Z} /2$ ) belonging to the fourth generation
necessarily implies an unnaturally  large hierarchy
among the Yukawa couplings of
different neutrinos with the Higgs boson . Though the naturalness
arguement has an intuitive appeal, it is basically a philosophical
outlook and,therefore, is not beyond debate. However
 the purpose of this note is to point out, without questioning the
validity of the philosophy of naturalness, that any
unnatural  hierarchy is not inevitable  for the existence of
such a heavy neutrino .On the
contrary the completely opposite scenario  \cite{harari,jarlskog,nambu}
referred to in the  literature as flavour democracy
, which requires that the Yukawa couplings of all fermions of a particular
type with the Higgs boson before the diagonalisation of the mass matrix
are equal, can be implemented in a four family  model
with a heavy neutrino without
requiring any large hierarchy among the Yukawa couplings of the fermions
\rm of\/{\it different types}.
 This is no longer possible in a three family
model  in view of the rather stringent lower  bound  on the top
quark  mass and the relatively low masses of the other
fermions (see below).

      The elements of the n x n mas matrix $M^{f}$ of the
fermions of the type f ( up, down, charged lepton or neutrino )
in an n--generation model with approximate flavour democracy can
be written as
\begin{equation}
M^{f} ~ = ~ Y^{f} ( M^{0} \; + \;\lambda  M_{f}^{\prime })
\end{equation}
where $ M^{0}$ is the fully democratic  matrix with all
elements equal to unity, $Y^{f}$ can be interpretated within the framework
of the MSM as the common Yukawa coupling of the  fermions of this
type  (in units of the vacuuam expectation value of the Higgs boson).
The parameter $\lambda$ is introduced purely for book keeping
purposes and we take it to be 0.1. The matrix $M^{\prime}$ with elements
$ O(1 )$  parametrises small departures from perfect democracy.
 In the limit
when $M^{\prime}$ vanishes this mass matrix can be motivated by imposing
a permutation symmetry \cite{harari} or by an underlying BCS like
dynamics \cite{nambu}. Any theoretical
idea which pinpionts the origin of flavour democracy  and the mechanisms
for small departures from it will, of course, be an important  step forward.
 In this phenomenological work focussed mainly on the {\it  naturalness }
  of the hierarchies in the fermion masses and mixing angles
 within the frame work of the MSM, we shall
not speculate about  the above points.
 It is
well-known that n -- 1 eigenvalues of $M^{0}$
are equal to zero while one is equal to n. This illustrates that
a large hierarchy among the masses of a given species  f does not
necessarily require a corresponding hierarchy among the elements
of the mass matrix. The problem of a three family model is
apparent once the inter-- species mass hierarchy is taken into account.
This model predicts :
\begin{equation}
m_{t}: m_{b}: m_{\tau}: m_{\nu} ~  =  ~ Y^{u}: Y^{d}: Y^{l}: Y^{\nu}
\end{equation}
where $m_{\nu}$ stands for the mass of the heaviest neutrino.
Using the known values of $m_{b}$ and $m_{\tau}$ and the bounds $m_{t}\; \geq $
91 GeV \cite{cdf} and $m_{\tau}\; \leq $ .035 GeV
\cite{pdata}, one requires $ Y^{u}: Y^{d}\geq $ 18 , $ Y^{u}: Y^{l}\geq$
 50 and $ Y^{u}: Y^{\nu}\geq $ 2570 ! While the first ratio is barely
consistent with the philosophy of  naturalness  the
second one is certainly not so. The third one, implying the largest
hierarchy, is a
reflection of the well-known neutrino mass problem .
 In contrast if nature indeed
prefers to have four families of quark--leptons with $m_{T} \simeq
m_{B} \simeq m_{E} \simeq m_{N}$ (the subscripts refer to fermions
belonging to the fourth family ), she may do it without hurting anybody's
cravings for naturalness.

     The non--zero masses of the lighter fermions within a given  type
f are generated by $M^{\prime}$ and at the first sight it seems
that there are too many paramers to fit the observed fermion spectrum
and the quark mixing angles. This ,however, is not the case. The
requirement that the departure from the democratic structure is
small i.e., the elements of $M^{\prime}$ are  $ O(1 )$ makes this model
quite restrictive.
In the following we shall illustrate with an extremely simple model
that the known phenomenology of the quarks and leptons can be
understood without destroying approximate democracy  or without resorting
to fine tuning. Since the mass hierarchy in the charged lepton sector
is not too different from the down quark sector a similar model
can be applied there. As long as there is no positive evidence for non-zero
masses for the lighter neutrinos a fully democratic mass matrix,
 rather than the omission of the right--handed neutrinos by hand,
seems to be more natural  for this sector. However, in order to
accommodate a small non--zero mass of  any one of them consistent
with the present upperbounds \cite{pdata}
some fine tuning is required ~( see below ) though no large ratio
of the $Y^{f}$'s needs to be introduced.
Should such a situation arise the simple interpretation of eq.(1)
as the roughly equal Yukawa couplings
of a single Higgs boson may appear to be
unattractive and physics beyond the MSM may be called for. We
will comment on it at the end of the paper.

   In order to construct an illustrative simple model we obtain the
approximate eigenvalues and eigenvectors of $M^{f}$ by applying
perturbation theory. Since the degenerate eigenvectors of $M^{0}$ can not be
determined  uniquely without specifying additional symmetries we
have used instead the matrix elements $(M^{f})_{\alpha \beta}$ , where
$\alpha$ , $\beta$ $~ = ~$ 1,2,3 label the degenerate eigenvectors
of $M^{0}$ while 4  refers to the eigenvector corresponding to the
heavy state,
as the phenomenological parameters for the subsequent analysis.It
should ,however be emphasised that the above ambiguity is an artifact
of using perturbation theory and no physical observable is  affected
by it. For simplicity we also take $M^{\prime}$ to be real
symmetric. It is well-known that CP violation in  a four generation
model with three observable phases  is less restrictive than in a
three generation model with a single observable phase\cite{datta}.
It is, therefore, not unreasonable to assume that the introdution
of three small phases will adequately describe CP violation without
drastically altering the pattern of quark masses and mixing angles
obtained here.

	   The fact that $m_{u},m_{c} << m_{t}$ gives a strong hint
regarding the input values of these matrix elements. It is natural
to assume that u and c remain massless and degenerate upto first
order in perturbation theory while the t quark picks up a mass.There
are two attractive ways of achieving this without adjusting the
detailed numerology of the matrix elements : i) Assume that all
$(M^{\prime})_{\alpha \beta}$  except $(M^{\prime}_{33})$ are zero
in the subspace of the degenerate eigevectors or ii)
$(M^{\prime})_{\alpha \beta}$  also has a democratic structure in this
subspace. The second alternative can be reduced to case i) by
diagonalising the perturbation matrix and can be analysed similarly.
Using assumption i)
we have ten remaining free parameters $(M_{q}^{\prime})_{33}$,
$(M_{q}^{\prime})_{44}$
and $(M_{q}^{\prime})_{i4}$ (i=1,2,3; q=u or d) which are to be determined
phenomenologically.

    Applying the standard tools of degenerate perturbation theory
it is now straight forward to see that one quark remains massless
to all orders  in perturbation theory,which can be identified
with the lightest quark of a given type (up or down). The 2 x 2 block
of the CKM matrix involving the first two families are generated
in the zeroth order. All other elements are $ O(\lambda )$ and
are smaller.
The relevant
formulae for the remaining masses are :

\begin{equation}
m_{q}^{(4)}  ~ = ~ Y^{q}(4 \; + \;\lambda(M_{q}^{\prime})_{44} )
\end{equation}
\begin{equation}
m_{q}^{(3)}  ~ = ~ \lambda Y^{q} (M_{q}^{\prime})_{33}
\end{equation}
\begin{equation}
m_{q}^{(2)}  ~ = ~ \frac{\lambda^{2} Y^{q}}{4} ((M_{q}^{\prime})_{14}^{2}
 \; + \;
(M_{q}^{\prime})_{24}^{2} )
\end{equation}
where q = u or d and the superscripts 2,3,4 respectively denotes c, t, T
(s, b, B ) quarks in the up ( down ) sector. The elements of the CKM
matrix , whose measured values are used to fix the remaining
free parameters of the model,are given by(upto $ O(\lambda)$ )
\begin{equation}
V_{ud}  ~ = ~ N_{u} N_{d} (1 \; + \; x_{u} x_{d} )
\end{equation}
\begin{equation}
V_{us}  ~ = ~ N_{u} N_{d} (x_{d} \; - \;   x_{u} )
\end{equation}
\begin{equation}
V_{ub}  ~ = ~ \eta_{ub} \frac{\lambda}{4 N_{d}} V_{us}
\frac{(M_{d}^{\prime})_{34} (M_{d}^{\prime})_{24}}{(M_{d}^{\prime})_{33}}
\end{equation}
\begin{equation}
V_{cb}  ~ = ~ \eta_{ub} \eta_{cb} N_{u} N_{d} \frac{|V_{ub}|}{V_{us}}
 (1 \; + \; x_{u} x_{d} \; - \; \eta_{ub} \frac{|m_{c}|}
 {m_{t}} \frac{(M_{u}^{\prime})_{34}}{(M_{u}^{\prime})_{24}}
 \frac{|V_{ub}|}{V_{us}} )
\end{equation}
where $N_{u,d} = ( 1 + x_{u,d}^{2})^{-1/2}$ , $x_{u,d} =
((M_{u,d}^{\prime})_{14} / (M_{u,d}^{\prime})_{24} )$ and
$\eta_{ub}$, $\eta_{cb} = \pm 1 $ refer to the signs of $V_{ub}$ and
$V_{cb}$. $(M_{u}^{\prime})_{14}$ and $(M_{d}^{\prime})_{24}$
also involves sign ambiguities. We choose them to be positive.
It is to be noted that eq.(6) holds to all orders in perturbation
theory while the remaining three receives higher order corrections.
We have computed upto second order corrections to the above formulae.
The expressions are  rather
cumbersome and will be presented in a longer paper
\cite{datta1}.The numerical values of these corrections are however
given below to demonstrate that perturbation theory makes sense.Finally
we have  rechecked  everything  by using numerical
diagonalisation without appealing to perturbation theory (see below).
In the absence of any experimental guidelines regarding $m_{T}$ and
$m_{B}$ we have to assume some
reasonable values for $Y^{u}$ and $Y^{d}$. We take
 $Y^{u}$ = 150, $Y^{d}$  = 100 . The parameters $(M_{u}^{\prime})_{44}$
 and $(M_{u}^{\prime})_{44}$ do not affect the  masses of the quarks
 belonging to the first three families
or their  mixing angles to the lowest order in $\lambda$ . We have chosen
$(M_{u}^{\prime})_{44} \simeq - (M_{d}^{\prime})_{44} \simeq$ - (
8.0--8.5). The last choice is made so that the
mass splitting of the fourth generation quarks do not make a large
contribution to the $\rho$ parameter \cite{veltman,foot}. We emphasize
that this choice is not necessary if one takes  $Y^{u} \simeq  Y^{d}$.
Our choices of $Y^{u}$ and $Y^{d}$  gives somewhat
better hierarchy between the up and the down sectors but is not crucially
important.
Using eqs. (3)--(8) we obtain ( for $m_{c} \simeq$ 1.5
$m_{t} \simeq$ 125, $m_{s} \simeq$ .150, $m_{b} \simeq$ 5 (all in
GeV)
, $V_{ud}\simeq$0.9747, $V_{us}\simeq$ 0.223, $V_{ub}\simeq$ 0.004
and $V_{cb}\simeq$0.04 ) : $x_{u} = 0.2$, $x_{d} = 0.45$,
$(M_{u}^{\prime})_{14}\simeq$ 0.392, $(M_{u}^{\prime})_{34}\simeq$3.76
, $(M_{u}^{\prime})_{33}\simeq$8.33, $(M_{d}^{\prime})_{14}\simeq$0.312
, $(M_{d}^{\prime})_{34}\simeq$-0.462, $(M_{d}^{\prime})_{33}\simeq$0.5.
It is gratifying to note that no large hierarchy in the matrix elements
is required.The remaining elements of the CKM matrix are
predictions of this model and the full matrix is given by
\begin{equation}
V =\left( \begin{array}{cccc}
0.9747(0.0) & 0.2235(-0.00008) &
0.004(-0.0009) & 0.004(-0.0009) \\
-0.2235(0.0003) & 0.9747(-0.0012) &
0.04(0.001) & -0.031(-0.017) \\
0.005(0.002) & -0.039(-0.004) & 1.00(-0.006)
& -0.105(-0.036)\\
-0.011(-0.002) & 0.03(0.01) & 0.105(0.037) &
1.00(-0.006) \\
\end{array} \right)
\end{equation}
\vskip 0.1in
where the numbers in the parentheses give the second order corrections.
While  the detailed numerology of the predictions of this  matrix is somewhat
dependent on the values of the input parameters, a key prediction
independent of such details is that  the 2 x 2 block involving
the first two families should be almost identical. The present
experimental values $|V_{cd}|$ = 0.204 $\pm$ 0.017 and
$|V_{cs}|$ = 1.00  $\pm$ 0.20 \cite{pdata} leave  ample room for deviations
from this prediction.
 The discovery of the fermions belonging to the
fourth family by direct searches in conjuction with accurate measurements
of $V_{cd}$ and $V_{cs}$  at a tau--charm factory may provide an
an interesting test of this model.

  Using the phenomenologically determined
$(M_{u,d}^{\prime})_{\alpha \beta}$'s and a specific choice for
the degenerate eigenvectors of $M^{0}$ it is an easy numerical
excercise to determine the elements of $M_{u,d}^{\prime}$. As has
already been mentioned the physical observales are of course
independent of this choice. As  example we present below two
such matrices .  One can readily
diagonalise them  numerically to verify that they indeed
reproduce all the known phenomenology of the quarks and a close
agreement with the CKM matrix given above.
\begin{equation}
M^{u} =\left( \begin{array}{cccc}
1.0775 & 0.9435 & 1.0497 & 0.5287\\
0.9435 & 0.8095 & 0.9158 & 0.3947\\
1.0497 & 0.9158 & 1.0220 & 0.5010\\
0.5287 & 0.3947 & 0.5010 & 1.0910\\
\end{array} \right)
\end{equation}
\vskip 0.1in
\begin{equation}
M^{d} =\left( \begin{array}{cccc}
1.2505 &  1.1960 & 1.2280 & 1.2348\\
1.1960 &  1.1415 & 1.1735 & 1.1803\\
1.2280 &  1.1735 & 1.2055 & 1.2124\\
1.2348 &  1.1803 & 1.2124 & 1.2858\\
\end{array} \right)
\end{equation}
\vskip 0.1in

As expected the departure from democracy is rather small in the
down sector. In the up sector the departure from democracy outside
the 3 x 3 block is quite significant. Whether this is an artifact
of our input parameters  has to be checked by a detailed numerical
computation\cite {datta1}. We also note that the above precise
values of the elements of the mass matrices are not required to
get the hierarchy in the masses qualitatively. These are required
to reproduce the rather precisely determined $V_{ud}$ and $V_{us}$.

As has already been mentioned ,  nonvanishing neutrino masses of a
few MeV or smaller may revive the  naturalness problem in this model.
For example, a tau--neutrino mass close to its present upper limit
would imply $\lambda Y^{\nu} (M_{\nu}^{\prime})_{33} \leq$
0.035 GeV or $(M_{\nu}^{\prime})_{33} \leq$ 0.003 (assuming $Y^{\nu}
\simeq$ 100.0). This would reintroduce  large hierarchies among the
matrix elements. In the context of the neutrinos it may, therefore, turn
out be more appealing to use $(M_{\nu}^{\prime})_{33} $ = 0.0 as
an  input. Such a model would lead to two massless neutrinos to
all orders in perturbation theory and a light neurino with mass
\begin{equation}
m_{\nu}^{(3)}  ~ = ~ \frac{\lambda^{2} Y^{\nu}}{4}((M_{\nu}^{\prime})_{14}
 \; + \;(M_{\nu}^{\prime})_{24}\; + \;(M_{\nu}^{\prime})_{34} )
\end{equation}
A tau neutrino with mass $\simeq$ 0.035 GeV  would then imply
(assuming approximate equality of all matrix elements )
$(M_{\nu}^{\prime})_{i4}\simeq$ 0.2, which does nor seem to be  unnatural.

	It is well-known that if all the neutrino masses  turn out to
be non-zero and  very small, they can be accommodated  in the three family
model only at the expense of introducting large hierarchies
in the inter--species Yukawa couplings.
In a four family scenario
no such lagre hierarchy needs to be introduced but fine-tuning
of the values of these couplings seems to be necessary which makes
one feel rather uneasy. Perhaps new physics with an inbuilt see--saw
mechanism \cite{gellmann} ( as in ref 3) together with family democracy
 will provide an elegant model. Nevertheless, since no
 evidence of the Majoraana nature of the neutrinos which is so crucial
for the see--saw mechanism has been found , it  may be
worhwhile to  speculate about
alternative scenarios within the context of the present model .
 If the quarks and leptons are composites
 of more fundamental objects such a scenario can be motivated.In
order to make the discussion as much  model dependent
as is possible we simply
summarise the essential features . We assume that the Yukawa couplings
of the (composite) fermions with the Higgs boson are fully democratic.
The compositeness of the fermions manifests it self through non-
renormalisable interactions of the form $ (g_{ij}^{2}/\Lambda^{2})
{\bar \psi_{iL}}  \psi_{jR} {\bar f_{iL}} f_{jR}$ where the $\psi$'s and f's
stand for
usual quark--leptons and preons respectively\cite{pati}, $g_{ij}$'s
are coupling
constants and $\Lambda$ is the compositeness scale. The formation
of preonic condensates through some new interactions then generates
additionnal mass terms for the fermions which lead to departures from
exact democracy.The parameter $\lambda Y^{f}$  in eq. (1) can then be
identified with $g_{ij}^{2} \Lambda_{f}^{3}/ \Lambda^{2}$, whwre
$\Lambda_{f}^{3}$ denotes the strengthes of the preonic condensates.
 If the  compositeness scale of
the neutrinos is much larger than the other fermions then the departures
from the democratic structure in the neutrino sector will
naturally be much
smaller than those in the other sectors. Assuming $\lambda Y^{f}\simeq$
10,  $g_{ij}\simeq$ 1
and assigning a typical value of 300 $GeV^{3}$  characteristic of the
electroweak scale to the preon condensates , one needs $\Lambda
\simeq$ 2 TeV in order to produce the above mass matrices in the
 quark sector. Such a low compositeness scale in
a one scale model ,though allowed by the present experimental
data , may lead to unacceptably large flavour changing neutral currents
\cite{lyon} unless special mechanisms are introduced to suppress them.
On the other hand in  two scale models \cite{pati} the above
effective interactions may be generated by the exchanges of a composite
scalar particle of mass in the TeV region characteristic of the lower scale.
In such models flavour changing neutral currents are
also  naturally suppressed.

	  In summary,  the excitement
that the neutrino counting experiments have also determined the number
of sequential quark lepton families seems to be
rather premature. In this note we have illustrated  with
the help of an extremely simple model that the hierarchy among
the  fermion masses at least in the quark and charged lepton
 sectors can be understood
naturally in models with (approximate) flavour  democracy provided
there are four sequential families  of quark--leptons
 with all members of
the fourth family having roughly equal masses. This is not
possible in a three family model.
 In the neutrino sector
exact flavour democracy predicts three mass less neutrinos without
requiring the absence of right-handed neutrinos.
Departures from flavour democracy can naturally explain one massive
neutrino belonging to the lighter families provided its  mass
is in the MeV region. If confronted
with non-vanishing but extremely small neutrino masses for the first
three families the model may still accommodate  them without
introducing any large ratios , although some fine tuning in the values
of  the Yukawa  couplings may be needed. The possibility of understanding
very light sequential Dirac neutrinos in the democratic scenario
with composite quark--leptons has been qualitatively discussed.

   The author  wishes  to thank Prof. Abdus Salam, IAEA and UNESCO for
supporting his  visit to the International Centre for
Theoretical Physics, Trieste where part of this work was done.
He thanks Dr.A.Joshipura for valueable discussions.
This research was partly supported by a grant from the Department
of Science and Technology ,India.

\end{document}